\documentclass[12pt]{report}

\usepackage[dvips]{graphicx}
\usepackage{amssymb}
\usepackage{amsmath}

\makeatletter
\renewcommand{\@oddhead}{\hfil \thepage}
\renewcommand{\@oddfoot}{}
\renewcommand{\thefootnote}{\alph{footnote}}

\makeatother

\begin{document}

\begin{center}
\textbf{\large {Highly Excited Friedmann Universe}} \footnote{$\ $
Published in \textbf{Physics of Atomic Nuclei}, Vol. 62, No. 9,
1999, pp. 1524-1529. Translated from \textbf{Yadernaya Fizika},
Vol. 62, No.9, 1999, pp. 1625-1631.}
\end{center}
\begin{center}
\textbf{V. V. Kuzmichev} \footnote{$\ $ e-mail:
vvkuzmichev@yahoo.com; specrada@bitp.kiev.ua}
\end{center}
\begin{center}
\textit{Bogolyubov Institute for Theoretical Physics, National
Academy of Sciences of Ukraine, Metrolohichna St. 14b, Kiev, 03143
Ukraine}
\end{center}
\setcounter{footnote}{0}
\renewcommand{\thefootnote}{\arabic{footnote}}

\textbf{Abstract:} A highly excited Friedmann universe filled with
a scalar field and radiation has been considered. On the basis of
a direct solution to the quantum-mechanical problem with a
well-defined time variable, it has been shown that such a universe
can have features (energy density, scale factor, Hubble constant,
density parameter, matter mass, equivalent number of baryons, age,
dimensions of large-scale fluctuations, amplitude of fluctuations
of cosmic microwave background radiation temperature) identical to
those of the currently observed Universe.\\[0.5cm]

\textbf{1.} Available cosmological data suggest that, from the
point of view of quantum theory, the currently observed Universe
is likely to be in a highly excited state [1]. This is confirmed
by estimates of the number of the quantum state that corresponds
to its averaged motion as a discrete unit [1-4]. In view of this,
it is important to investigate cosmological systems featuring
gravitational and matter fields and occurring in states with large
quantum numbers by proceeding from a direct solution to the
relevant quantum-mechanical problem.

A model of the Friedmann universe filled with a uniform scalar
field has been proposed in [4]. This model, which is appropriate
for constructing a quantum theory, features a well-defined time
variable. The reference frame was specified there with the aid of
a subsidiary matter source in the form of radiation (relativistic
matter of any nature) that was assumed to be initially present in
the cosmological system along with a scalar field that forms a
nonzero cosmological constant in the early universe. The evolution
of the universe filled with not only scalar field but also with
radiation differs from that which is realized in the absence of
radiation. The main difference lies here in the emergence of a new
region that is accessible to a classical motion and which is
bounded by the potential barrier existing in the system of scalar
and gravitational fields. A quantum universe involving a slowly
varying scalar field and occurring in low-lying (quasistationary)
states has been analyzed in [4], where it has been shown that the
dynamical model proposed there is compatible with the currently
prevailing ideas of the early Universe. In this study, we will
consider a quantum Friedmann universe with large quantum numbers
characterizing the possible physical states of the gravitational
and matter fields involved. On the basis of a solution to the
quantum mechanical problem, it is shown that the universe in
highly excited states can have features (energy density, scale
factor, Hubble constant, density parameter, matter mass,
equivalent number of baryons) identical to those in the currently
observed Universe. The problem of the age of the Universe and the
possible new mechanism that could generate fluctuations of the
metric due to a finite width of quasistationary states and to the
anisotropy of cosmic microwave background radiation are discussed
in the Appendices.

\textbf{2.} The wave function of the quantum Friedmann universe
filled with radiation and a uniform scalar field $\phi $ specified
by the potential $V(\phi )$ is determined by the Schr\"{o}dinger
equation [4]
\begin{equation}
  2\,i\,\partial _{T} \Psi = \left( \partial _{a}^{2} -
      \frac{2}{a^{2}}\,\partial _{\phi }^{2} - U \right)  \Psi ,
\label{1}
\end{equation}
where $T$ is a privileged time coordinate related to the
synchronous proper time $t$ by the coordinate condition $dT / dt =
1 / a$, $a$ being the scale factor, while
\begin{equation}
          U = a^{2} - a^{4}\,V(\phi )
\label{2}
\end{equation}
plays the role of the effective interaction potential in the
cosmological system being considered. Here, we use the system of
units in which $l = \sqrt{2 G \hbar / 3 \pi c^{3}} = 1 $ and
$\tilde \phi = \sqrt{3 c^{4} / 8 \pi  G} = 1.$\footnote{$\ $In
order to go over to the system of units where $\hbar = c = 1$, we
must use the relation $l \tilde \phi ^{2} = \sqrt{3 / 32 \pi
^{3}}\,m_{p}$ for the energy and the relation $(\tilde \phi / l
)^{2} = (9 / 16)\,m_{p}^{4}$ for the energy density, $\ m_{p}$
being the Planck mass.}

Possible solutions to equation (1) are determined by the
properties of the scalar field involved. If $V(\phi )$ is
everywhere positive definite, we can see that, in the variable
$a$, the potential $U$ has the form of a barrier with height
$U_{max} = \frac{1}{4 V}$ and width $\Delta a = a_{2}(E) -
a_{1}(E)$, where $a_{1}(E)$ and $a_{2}(E)$, $a_{1}(E) < a_{2}(E)$,
are the classical turning points determined from the condition $U
= E$. With the aid of the equation of motion for the field $\phi
$, it can be shown that, if the field satisfies the condition $
|\partial _{t}^{2} \phi | \ll |\frac{dV}{ d\phi}|$, the
contribution of the operator $\frac{2}{a^{2}} \partial _{\phi
}^{2}$ can be approximated by the term $-\, \frac{a^{4}}{18}
\left( \frac{1}{H} \frac{dV}{d\phi }\right)^{2}$, where $H =
\frac{\partial _{t} a}{a}$ is the Hubble constant; that is, this
contribution is equivalent to the squared addition to the
interaction $U$ of the gradient of the potential $V(\phi )$. In
this effective potential, stationary states cannot exist in the
region $a \leq a_{1}$. If, however, $V(\phi ) \ll 1$,
quasistationary states with lifetimes exceeding the Planck time
can exist within the barrier [4]. The positions and widths of such
states are determined by the solutions to equation (1) that
satisfy the boundary condition in the form of a wave traveling
toward greater values of $a$.

A general solution to equation (1) can be represented in the integral form
\begin{equation}
  \Psi (a, \phi , T) = \int_{0}^{\infty }\!dE\,\mbox{e}^{\frac{i}{2} E T}\,
               C(E)\, \psi _{E}(a, \phi ),
\label{3}
\end{equation}
where the function $C(E)$ characterizes the $E$ distribution of
the states of the universe at the instant $T = 0$, while $\psi
_{E}(a, \phi )$ and $E$ are, respectively, the eigenfunctions and
the eigenvalues for the equation
\begin{eqnarray}
 \left( -\,\partial _{a}^{2} + \frac{2}{a^{2}}\,\partial _{\phi }^{2} +
             U - E  \right)  \psi _{E} = 0.
\label{4}
\end{eqnarray}

\textbf{3.} Let us now consider a quantum universe where
$\left|\frac{1}{V} \frac{dV}{d\phi }\right|^{2} \ll 1$. A solution
to equation (4) can then be represented as
\begin{eqnarray}
    \psi _{E}(a, \phi ) = \int_{0}^{\infty }\! d\epsilon \,
    \varphi _{\epsilon }(a, \phi )\, f_{\epsilon}(\phi ; E),
\label{5}
\end{eqnarray}
where $\varphi _{\epsilon }$ and $\epsilon $ are, respectively,
the eigenfunctions and eigenvalues for the operator $\left[-\,
\partial _{a}^{2} + U \right]$ of the adiabatic
approximation that correspond to continuum states at a fixed value
of the field $\phi $. The functions $\varphi _{\epsilon }$ can be
normalized to the delta function $\delta (\epsilon  - \epsilon
')$. Their form greatly depends on the value of $\epsilon $. For
$\epsilon < U_{max}$, there are quasistationary states with
$\epsilon = \tilde {\epsilon }_{n} \equiv \epsilon _{n} +
i\,\Gamma _{n}$ in the system [4], and the main contribution to
the integral in (5) over the interval $0 < \epsilon < U_{max}$
comes from the values $\epsilon \approx \epsilon _{n}$ and $a <
R$, where $R \geq a_{2}(\epsilon )$. Here, the wave function has
the form
\begin{equation}
  \varphi _{\epsilon } = A(\epsilon )\,\varphi _{\epsilon }^{(0)},
\label{6}
\end{equation}
where the function $A(\epsilon )$ has a pole in the complex plane
of $\epsilon $ at $\epsilon = \tilde {\epsilon _{n}}$, while
$\varphi _{\epsilon }^{(0)}$ is the solution over the interval $0
< a < R$ that is regular at the point $a = 0$,  normalized to
unity, and  weakly dependent on $\epsilon $. Proceeding in a way
similar to that adopted in the theory of quasistationary states
for short-range potentials [5], we can show that the quantity
$|A(\epsilon )|^{2}$ can be approximated by the delta function
$\delta (\epsilon  - \epsilon _{n})$ in the case of the potential
(2) as well. In this approximation, expression (5) can eventually
be reduced to the expansion
\begin{equation}
   \psi _{E}(a, \phi ) = \sum_{n}\,\varphi _{n}(a, \phi )\,
      f_{n}(\phi ; E) + \int_{U_{max}}^{\infty }\! d\epsilon \,
    \varphi _{\epsilon }(a, \phi )\, f_{\epsilon}(\phi ; E),
\label{7}
\end{equation}
where $\varphi _{n} = \varphi _{\epsilon _{n}}^{(0)}$ for $0 < a <
R$ and $\varphi _{n} = 0$ for $a > R$ (a state of this type was
considered in [6]), while $f_{n}(\phi ; E) = \int_{0}^{\infty
}\!da\,\varphi _{n}^{*}(a, \phi )\, \psi _{E}(a, \phi )$. In the
limit of an impenetrable barrier, the function $\varphi _{n}$
reduces to the wave function of a stationary state with a definite
value of $\epsilon _{n}$. In the case of $V \ll 1$ considered
here, the contribution of the integral to the expansion in (7) can
be disregarded, and the quantities $f_{n}(\phi ; E)$ can be
interpreted as the amplitudes of the probability that the universe
is in the state $f_{n}(\phi ; E)$ with a given value of the field
$\phi $. They satisfy the set of differential equations
\begin{equation}
 \partial _{\phi }^{2} f_{n} +
 \sum_{n'} K_{n n'} \left( \phi ; E \right) f_{n'} = 0,
\label{8}
\end{equation}
where
\begin{eqnarray}
 K_{n n'} =
  \langle \varphi _{n} | \partial _{\phi }^{2} | \varphi _{n'} \rangle +
  2\,\langle \varphi _{n} | \partial _{\phi } | \varphi _{n'} \rangle
  \partial _{\phi } +  \frac{1}{2}\,
  \langle \varphi _{n} | a^{2} | \varphi _{n'} \rangle
  \left( \epsilon _{n'} - E \right).
\label{9}
\end{eqnarray}
Over the time interval $\Delta T < \frac{1}{\Gamma _{n}}$, we can
disregard the possibility of decay and consider the
quasistationary state as a stationary state that arises in place
of the quasistationary state when the decay probability tends to
zero. The wave function $\varphi _{n}$ of such a stationary state
and the corresponding eigenvalue $\epsilon _{n}$ can be found by
perturbation theory by considering the interaction $a^{4} V(\phi
)$ as a small perturbation against $a^{2}$ (in the region $a <
a_{1}$, we have $a^{2} V < 1$). This yields
\begin{eqnarray}
          \varphi _{n}  & = & | n \rangle -  \frac{V}{4}\,\left[
        \frac{1}{8}\, \sqrt{N(N - 1)(N - 2)(N - 3)}\,| n - 2 \rangle +
                                          \right.             \nonumber\\
                        & + &
   \,\sqrt{N(N - 1)}\left(N - \frac{1}{2}\right) | n - 1 \rangle -
                                                  \nonumber\\
                        & - &
    \,\sqrt{(N + 1)(N + 2)}\left(N + \frac{3}{2}\right) | n + 1 \rangle -
                                                  \nonumber\\
                        & - &
    \,\left. \frac{1}{8}\, \sqrt{(N + 1)(N + 2)(N + 3)(N + 4)}\,
          | n + 2 \rangle \right] + O\left(V^{2}\right),
\label{10}
\end{eqnarray}
\begin{eqnarray}
          \epsilon _{n} & = & \epsilon _{n}^{0} - \frac{3}{4}\,V\,\left[ 2N(N + 1) + 1 \right] -
                                                        \nonumber\\
                        & - &
    \frac{V^{2}}{4}\,\left[\frac{17}{2}\,N^{3} + \frac{51}{4}\,N^{2} +
   \frac{59}{4}\,N + \frac{21}{4}\right] + O\left(V^{3}\right),
\label{11}
\end{eqnarray}
where $N = 2\,n + 1$, while $\epsilon _{n}^{0} = 2N + 1$ and $| n
\rangle$ are, respectively, an eigenvalue and the corresponding
eigenfunction of the operator $\left[-\, \partial _{a}^{2} + a^{2}
\right]$. Quasistationary states are realized for $V < 0.08 = 4.5
\times 10^{-2} m_{p}^{4}$ and are characterized by values
$\epsilon _{n} > 2.6$ and $\Gamma _{n} \ll 0.3$ [4]. The universe
can undergo a tunnel transition to the region $a > a_{2}(\epsilon
_{n})$ from any quasistationary state.

In the early universe, the quantity $V(\phi )$ specifies the
vacuum energy density, which determines the cosmological constant
$\Lambda $ at that era [7, 8]. In our Universe, the cosmological
constant is very small. In the model featuring matter in the form
of one scalar field, the reduction of the cosmological term can be
described in terms of the potential $V(\phi (t))$, which decreases
with time [3, 9]. A similar behavior of $V(\phi (t))$ is suggested
by the results of investigations within inflationary models [8,
9]. It was shown in [4] that, with a nonzero probability, a
quantum universe initially filled with radiation and a scalar
field whose potential $V(\phi (t))$ decreases with time can evolve
in the region bounded by the barrier. The expansion occurs here
owing to transitions from lower to higher states via the
interactions of the scalar and gravitational fields involved. In
the zeroth approximation, the evolution of the universe can be
described by considering transitions between unperturbed states $|
n \rangle $ and $| n + 1 \rangle $ under the effect of the
interaction $a^{4} V$ [4]. In a more rigorous approach that takes
into account variations in $V(\phi )$, the universe expands
through transitions between the states $\varphi _{n}$ and $\varphi
_{n + 1}$ due to the gradient of the potential $V(\phi )$.

As the potential $V(\phi )$ decreases, the number of quantum
states in which the universe can occur increases. As a result, the
competition between the tunneling processes and transitions from
one state to another arises in the universe that has not had time
to undergo tunneling. Since the decay probability decreases
exponentially with decreasing $V$, the probability that the
universe is excited to states with large quantum numbers $n$ in
the course of time is nonzero. The probability that, in course of
time, the universe in large-$n$ states will occur in the region
outside the barrier is negligibly small, because small values of
$V$ correspond to such states. In the limit $V \rightarrow 0$, the
universe is completely locked in the region within the barrier
with $\epsilon _{n}$ tending $\epsilon _{n}^{0}$.

\textbf{4.} Let us investigate the properties of the universe in
states with $n \gg 1$. In such states, $|V| \ll 1$ and $\Gamma
_{n} \sim 0$, and we can use the wave function (10) in calculating
the quantities $K_{n n'}$ in (9). As before, we assume that the
potential $V(\phi )$ changes slowly in these states as well
($\left|\frac{dV}{d\phi }\right| \ll 1$). The derivatives
$\partial _{\phi }\,\varphi _{n}$ in (9) can then be disregarded.
Taking this into account and going over to the limit $n \gg 1$, we
can then reduce equation (8) to the simplified form
\begin{equation}
 \partial _{\phi }^{2} f_{n} +
           \omega _{n}^{2}( \phi ; E) f_{n} = 0,
\label{12}
\end{equation}
where
\begin{equation}
   \omega _{n}^{2}( \phi ; E) = 2 N^{2} \left[ 1 -
              \frac{E}{2 N} - 2 N V(\phi )\right].
\label{13}
\end{equation}
The potential of the field $\phi $ will be chosen in the form
$V(\phi ) = \frac{m^{2}}{2} \phi ^{2}$. From the condition
 $|V|
\ll 1$, it follows that the mass of the fields must be constrained
by the condition $m \ll \frac{m_{p}^{2}}{|\phi |}$. In this case,
only the region $|\phi | \ll m_{p}$ corresponds to extremely large
masses $m \geq m_{p}$. For masses satisfying the inequality $m \ll
m_{p}$, the field $\phi $ can be determined over the entire
interval $0 \leq |\phi | < \infty $; in this case, equation (12)
reduces to the Schr\"{o}dinger equation for a harmonic oscillator.
Under such conditions, the field $\phi $ can oscillate near the
minimum of the potential $V(\phi )$, causing the production of
particles.\footnote{$\ $ At $E = 0$, a similar mechanism leads to
the production of particles by the inflaton field, which is
identified with the scalar field $\phi $ [8]. In the case
considered here, the field $\phi $ for $n \gg 1$ states can be
treated as an effective field obtained upon averaging over the
internal degrees of freedom of real physical fields.} Substituting
the explicit expression for $V(\phi )$ into (13) and introducing
the new variable $x = (2 m^{2} N^{3})^{1/4} \phi $, we arrive at
the equation
\begin{equation}
     \partial _{x}^{2} f_{n} + (z - x^{2}) f_{n} = 0,
\label{14}
\end{equation}
where $z = \frac{\sqrt{2 N}}{m}\left(1 - \frac{E}{2 N}\right)$.
Equation (14) has solutions at $z = 2 s + 1 $, where $s =
0,1,2,\ldots $. Suppose that the universe occurs in a state with a
definite quantum number $s$. The function $f_{n}$ can then be
represented as
\begin{equation}
   f_{n s} = \frac{1}{\sqrt{s!}} \left(B_{n}^{\dagger}\right)^{s} f_{n 0},
   \quad B_{n} f_{n 0} = 0, \quad
    f_{n 0} = \left(\frac{1}{\pi }\right)^{1/4} \mbox{e}^{- x^{2} / 2},
\label{15}
\end{equation}
where $f_{n 0}$ is the state vector of the universe in the $n$th
state at $s = 0$, while $B_{n}^{\dagger} = \frac{1}{\sqrt{2}}(x -
\partial _{x})$ and $B_{n} = \frac{1}{\sqrt{2}}(x + \partial _{x})$ are
the operators that, respectively, create and annihilate particles
in this state and which satisfy the conventional commutation
relations $[B_{n},B_{n}^{\dagger}] = 1$ and $[B_{n},B_{n}] =
[B_{n}^{\dagger},B_{n}^{\dagger}] = 0$. Since $n \gg 1$, small
changes in $n$ do not affect the physical state of the universe
where there are $s$ particles.

The condition of quantization of $E$ has the form
\begin{equation}
    E = 2 N - (2 N)^{1/2} (2 s + 1) m.
\label{16}
\end{equation}
For $m \ll m_{p}$ and small $s$, the presence of the field $\phi $
has no effect on the properties of the universe because, in this
case, the approximate equality $E \approx 2 N$ holds to a high
precision, so that the universe is dominated by radiation. A
transition from the radiation-dominated universe to the universe
where matter (in the form of particles produced by the field $\phi
$) prevails occurs when, owing to an increase in the number of
particles, the second term in (16) becomes commensurate with the
first one. Let us consider the case where there are many particles
in a highly excited $n$th state of the universe. For $n \gg 1$ and
$s \gg 1$, the wave function has the form
\begin{equation}
    \psi _{n s}(a,\phi ) = \varphi _{n}(a) f_{n s}(\phi ),
\label{17}
\end{equation}
where
\begin{equation}
    \varphi _{n }(a) = \left(\frac{2}{N}\right)^{1/4}
         \cos\left(\sqrt{2 N}\,a - \frac{N \pi }{2}\right),
\label{18}
\end{equation}
\begin{equation}
    f_{n s}(\phi ) = \left(\frac{m (2 N)^{3/2}}{4 s}\right)^{1/4}
         \cos\left(\sqrt{2 s}\,(2 m^{2} N^{3})^{1/4}\,\phi  -
              \frac{s \pi }{2}\right).
\label{19}
\end{equation}

This wave function corresponds to the semiclassical solution to
equation (4) and is normalized to unity with allowance for the
fact that the probability of finding the universe in the region $a
> a_{2}$ is negligibly small.

Let us now calculate the energy density $\rho _{tot}$ for matter
and radiation in the universe described by the wave function (17).
The energy density for a classical field $\phi $ and radiation is
determined by the Einstein equation for the
$\left(^{0}_{0}\right)$ component and is given by [4, 10]
\begin{equation}
    \rho  = \frac{2}{a^{6}} \pi _{\phi }^{2} + V + \frac{E}{a^{4}},
\label{20}
\end{equation}
where $\pi _{\phi }$ is the momentum canonically conjugate to the
variable $\phi $. Assuming that all the quantities in (20) are
operator-valued, we set $\rho _{tot} = \langle \rho \rangle $,
where averaging is performed with the wave function (17).
Disregarding the variances $\langle a^{2}\rangle  - \langle
a\rangle^{2}$ and $\langle a^{6}\rangle  - \langle a\rangle^{6}$,
we obtain
\begin{equation}
    \rho _{tot} = \frac{2}{\langle a \rangle ^{6}}
       \langle \pi _{\phi }^{2}\rangle  + \langle V \rangle +
           \frac{E}{\langle a \rangle ^{4}}.
\label{21}
\end{equation}
Calculating the expectation values in (21), we arrive at the total
energy density in the universe in the form of the sum of the
energy densities for matter and radiation as in the general theory
of relativity [10]:
\begin{equation}
    \rho _{tot} = \frac{193}{12}\,\frac{M_{\phi }}{\langle a \rangle ^{3}}
         +  \frac{E}{\langle a \rangle ^{4}}.
\label{22}
\end{equation}
Here, $M_{\phi } = m s$, and $\langle a \rangle =
\sqrt{\frac{N}{2}}$ is the scale factor in the universe occurring
in an $n \gg 1$ state. The constant $E$ and the total mass
$M_{\phi }$ are related by the equation
\begin{equation}
   E = 4 \langle a \rangle \left[ \langle a \rangle - M_{\phi } \right].
\label{23}
\end{equation}
If
\begin{equation}
    \langle a \rangle = M_{\phi },
\label{24}
\end{equation}
there is no radiation; we then have $\rho _{tot} = \rho _{sub}$
and
\begin{equation}
   \langle a \rangle  =
        \left( \frac{193}{12}\,\frac{1}{ \rho _{sub}} \right)^{1/2}.
\label{25}
\end{equation}

\textbf{5.} Suppose that the system being considered occurs in an
$n \gg 1, \ s \gg 1$ state and that it is characterized by the
scale-factor value of $\langle a \rangle \sim 10^{28} \mbox{cm}$.
The condition in (24) will then be satisfied at $M_{\phi } \sim
10^{56} \mbox{g}$. This value coincides with the mass of matter in
the observed part of our Universe [10]. On the other hand, we set
$\frac{E}{\langle a \rangle ^{4}}$ to the density of cosmic
microwave background radiation energy in the present era,  $\rho
_{\gamma }^{0} = 2.6 \times 10^{- 10}$ GeV/cm$^{3}$ [11]. From
(23), we then find that, for $\langle a \rangle = 10^{28}
\mbox{cm}$, we have $\frac{M_{\phi }}{\langle a \rangle } = 1 -
0.7 \times 10^{- 5}$; that is, the equality in (24) holds to a
high precision in this case as well. We note that, although the
absolute value of $E$ is not small in the present era ($E \sim
10^{117}$ [4]), it is small in relation to $\langle a \rangle ^{2}
\sim 10^{122}$ (all estimates are presented here in the system of
units where $l = \tilde \phi = 1$) and can be disregarded. It is
the ratio of these quantities that determines the accuracy to
which relation (24) holds.

The current values of the scale factor ($\langle a \rangle \sim
a_{0} \sim  10^{28} \mbox{cm}$) and of the mean energy density
($\rho _{sub} \sim \rho _{0} \sim 10^{- 5}$ GeV/cm$^{3}$) satisfy
relation (25). Associating the well-known equality $a_{0} =
\left(\frac{\Omega _{0}}{\Omega _{0} - 1} \frac{1}{\rho
_{0}}\right)^{1/2}$
 with relation (25), we find that the effective
value of the density parameter is $\Omega _{0} = 1.066$; that is,
the geometry of the universe with the above features is close to
Euclidean geometry. It is characterized by the quantum-number
values of $n \sim \langle a \rangle ^{2} \sim 10^{122}$ and $s
\sim \frac{\langle a \rangle }{m}$. Taking the proton mass for
$m$, we obtain $s \sim 10^{80}$. The above value of $n$ complies
with existing estimates for our Universe (see [1-4]), while $s$ is
equal to the equivalent number of baryons [10]. The value found
for $\Omega _{0}$ corresponds to the Hubble constant value of
$H_{0}^{theory} \simeq 94$ km/(s Mpc). This value of Hubble
constant lies within the limits of experimental uncertainty of the
value of $H_{0}^{exp} = 80 \pm 17$ km/(s Mpc), which was obtained
with the aid of the Hubble cosmic telescope [12].

The approach developed here makes it possible to obtain realistic
estimates for the age of the Universe (see Appendix 1), for the
proper dimensions of the nonhomogeneities of matter which are
consistent with those of the observed large-scale structure of the
Universe, and for the amplitude of the fluctuations of the cosmic
microwave background radiation temperature. The resulting estimate
for the last quantity is close to the value extracted from ÑÎÂÅ
experimental data (see Appendix 2).

\textbf{6.} The above numerical estimates of the parameters of the
universe are of an illustrative character. If we nevertheless
associate them with our Universe, it can be concluded that the
values observed in the present era for the scale factor, the mass
of matter in the Universe, and the density of the cosmic microwave
background radiation energy satisfy relation (23). The zero-order
approximation, which corresponds to setting $\pi _{\phi }^{2} = 0$
in (20), leads to a very small value of $\Omega _{0} \simeq 0.1$,
but it is very close to the lower boundary of the uncertainty
interval for $\Omega _{0}$ [11]. Although the potential $V(\phi )$
undergoes only small variations in response to changes in the
field $\phi $, the field $\phi $ itself changes fast, oscillating
about the point $\phi  = 0$, so that the approximation in which
$\pi _{\phi }^{2} = 0$ is invalid. The application of the present
model in this approximation would result in the
radiation-dominated universe; that is, it would not feature a
mechanism capable of filling it with matter upon a slow descent of
the potential $V(\phi )$ to the equilibrium position, which
corresponds to the true vacuum. In states of the universe that are
characterized by large values of the quantum numbers, the kinetic
term of the scalar field ensures the density parameter value close
to unity.

Replacement of the entire set of actually existing massive fields
by some averaged massive scalar field seems physically justified
for states of the universe that have large values of $n$. We can
see that, by and large, such an averaged field describes correctly
the global features of our Universe. It effectively includes
visible baryon matter and dark matter, which is globally
manifested on cosmological scales via gravitational interaction.
The status of the field $\phi $ changes as we go over from one
stage of universe evolution to another. In the early universe, the
field $\phi $ ensures a nonzero value of the vacuum-energy density
(cosmological constant) due to $V(\phi )$ values at which the
equation for $\varphi _{\epsilon }(a, \phi )$ admits nontrivial
solutions in the form of quasistationary states. In a later era,
when the field $\phi $ descends to a minimum of the potential
$V(\phi )$ and begins to oscillate about this minimum, it appears
to be a source of some averaged matter filling the visible volume
of the universe, which has linear dimensions on the order of $\sim
\langle a \rangle $.

\begin{center}
APPENDIX 1
\end{center}

The time $\frac{1}{H_{0}}$ corresponding to the theoretical Hubble
constant value $H_{0}^{theory}$ is equal to $10^{10}$ yr, and the
age of the Universe calculated by the standard formula $t =
\frac{2}{3 H}$ is $t_{0} \approx 7 \times 10^{9}$
 yr. This result
is smaller than the expected value of $t_{0} = (10 \div 20) \times
10^{9}$ yr [11], but it is close to that which corresponds to
$H_{0}^{exp},\ t_{0} \approx 8 \times 10^{9}$ yr, highlighting the
problem of the age of the Universe. It should be borne in mind,
however, that the age of the universe is calculated by the formula
$t = \frac{q}{H}$, where $q = \frac{1}{2}$ (for the equation of
state $p = \frac{\rho }{3}$ or $\frac{2}{3}$ (for $p = 0$), which
implies that $a = b\,t^{q}$, where the proportionality factor $b$
is independent of $t$. In the model of the universe filled with
matter [as represented by the field $\phi (t)$] and radiation, the
factor $b$ depends on the variable $\phi (t)$, which changes with
time. The inclusion of this dependence leads to a greater value of
$t_{0}$. By way of example, we indicate that, for $2 \sqrt{V} t
\ll 1$, the scale factor varies with t according to the law $a(t)
\simeq (2 \sqrt{\epsilon }\,t)^{1/2}$, where $\epsilon = \epsilon
(\phi (t))$ [4]. From the above, it follows that, in order to
calculate the age of the universe, we must consider the
transcendental equation $t = \left[ 2 H - \partial_{t} \ln
\epsilon \right]^{- 1}$. Since $\epsilon > 0$ and since it
increases with time, we have $\epsilon > 0$.

In order to perform a numerical estimation, we assume that, in the
time interval being considered, $\epsilon $ grows in proportion to
a power of time: $\epsilon  \sim t^{\alpha }$. The age of the
universe is then given by the expression $t = \frac{1 + \alpha }{2
H}$. Since we have $\epsilon  \sim  10^{117}$ and $t \sim 10^{61}$
(in $l/c$ units), the exponent $\alpha $ is $\alpha \simeq 1.9$ [a
similar power-law dependence follows from the relation $a^{2} \sim
n \sim \epsilon $ and from the classical expression for $a(t)$],
whence we find that, if the Hubble constant set to its theoretical
value $H_{0}^{theory}$, the age of the universe is $t_{0} \simeq
15.2 \times 10^{9}$ yr. It should be noted that the above
power-law dependence of $\epsilon $ on $t$ ensures a correct
relationship between the current value of the scale factor $a_{0}$
and the age $t_{0}$ of the universe: $a_{0} \sim t_{0} \sim
10^{61}$.

\begin{center}
APPENDIX 2
\end{center}

A quasistationary state with a small, but finite value of the
width $\Gamma $ does not possess a definite value of $\epsilon $.
This uncertainty — we denote it by $\delta \epsilon $ — can serve
as source of fluctuations of the metric. Let us demonstrate this
explicitly. By associating  $\epsilon  + \delta \epsilon $ with
the scale factor $a + \delta a$ and by using the solution to the
Einstein equation in the region $a \leq a_{1}$ from [4], we find
that the amplitude of fluctuations of the scale factor can be
represented as
 $$
    \frac{\delta a}{a} = \frac{1}{4}\,
    \frac{\delta \epsilon / \epsilon}{1 - \mbox{tanh}\sqrt{V}t /2
    \sqrt{V \epsilon }}\ .
\eqno (A.1)
 $$
Since $\delta \epsilon \lesssim \Gamma $, the fluctuations $\delta
a$
 that were generated at the early stage of the evolution of the
Universe will take the greatest values. In order to estimate them,
we adopt the values of $V = 0.08,\ \epsilon = 2.6$, and $\delta
\epsilon \lesssim 0.3$, corresponding to the time $t \sim 1$ [4].
We then have $\frac{\delta a}{a} \lesssim 0.04$. Since the
dimension of large-scale fluctuations changed in direct proportion
$a(t)$, this relation has remained valid up to the present time
[13]. Taking this into account, we find that $\delta a \lesssim
130$ Mpc for the current value of $a \sim 10^{28}$ cm. On the
order of magnitude, the above value corresponds to the scale of
superclusters of galaxies [9]. Smaller values of $\delta \epsilon
$ are peculiar to quantum states with smaller $V$. The
fluctuations $\delta a $ corresponding to them are smaller than
those presented above and are expected to manifest themselves
against the background of the large-scale structure. They can be
associated with clusters of galaxies, galaxies themselves, and
clusters of stars. Thus, the conclusions drawn on the basis of the
model considered here are in line with the generally accepted
concept that galaxies, their clusters, and other structures in the
Universe are macroscopic manifestations of quantum fluctuations
that have grown considerably [9].

Let us estimate the amplitude of fluctuations of the cosmic
microwave background radiation temperature, $\delta T / T$. By
using the relation $\epsilon  = \rho _{\gamma }\, a^{4}$, where
$\rho _{\gamma } = (\pi ^{2} / 15)\, T^{4}$ is the density of the
cosmic microwave background radiation energy, we obtain
 $$
 \frac{\delta T}{T} = \frac{1}{4}\,
    \frac{\delta \epsilon}{\epsilon} - \frac{\delta a}{a}\ .
\eqno (A.2)
 $$
For  $\sqrt{V}\,t \ll 1$, it follows from (A.1) and (A.2) that
 $$
  \frac{\delta T}{T} \simeq  \frac{t}{2 \sqrt{\epsilon }}\,
     \frac{\delta a}{a}\ .
\eqno (A.3)
 $$
For the time $t \sim 10^{5}$ yr ( $t \sim 10^{56}$ in $l/c$
units), which corresponds to the recombination of primary plasma
(separation of radiation from matter), it can be found that, for
the observed value of $\epsilon = 5.2 \times 10^{117}$ (in which
case $\sqrt{V}\,t \leq \frac{t}{2 \sqrt{\epsilon }} \sim 0.7
\times 10^{- 3}$), the sought amplitude of fluctuations of cosmic
microwave background temperature radiation at $\frac{\delta a}{a}
\lesssim 0.04$ can be estimated as
 $$
  \frac{\delta T}{T} \lesssim  2.8 \times 10^{- 5}.
\eqno (A.4)
 $$
Upon recombination, the fluctuations of the temperature undergo no
changes; therefore, measurement of the quantity $\delta T / T$ for
the present era furnishes information about the Universe at the
instant of last interaction of radiation with matter. The estimate
in (A.4) is in good agreement with experimental data on cosmic
microwave background radiation (see [14]).

\begin{center}
REFERENCES
\end{center}

1. Hartle, J.B. and Hawking, S.W., Phys. Rev. D: Part. Fields, 1983,
vol. 28, p. 2960.

2. Zeidovich, Ya.B. and Novikov, I.D., Stroenie i evolyutsiya
Vselennoi (Structure and Evolution of the Universe), Moscow:
Nauka, 1975.

3. Kuzmichev, V.V., Yad. Fiz., 1997, vol. 60, p. 1707
[Phys. At. Nucl. (Engl. Transl.), vol. 60, p. 1558].

4. Kuzmichev, V.V., Yad. Fiz., 1999, vol. 62, p. 758 [Phys. At.
Nucl. (Engl. Transl.), vol. 62, p. 708].

5. Baz', A.I., Zel'dovich, Ya.B., and Perelomov, A.M., Scattering,
Reactions, and Decays in Nonrelativistic Quantum Mechanics,
Jerusalem: Israel Program of Sci. Transl., 1966.

6. Fock, V.A., The Principles of Quantum Mechanics, Mir: Moscow,
1978.

7. Al'tshuler, A.O. and Barvinskii, A.O., Usp. Fiz. Nauk, 1996,
vol. 166, p. 459.

8. Linde, A.D., Elementary Particles and Inflationary Cosmology,
Chur: Harwood, 1990.

9. Dolgov, A.D., Zeldovich, Ya.B., and Sazhin, M.V, Kosmologiya
rannei vselennoi (Cosmology of the Early Universe), Moscow: Mosk.
Gos. Univ., 1988.

10. Misner, C.W., Thorne, K.S., and Wheeler, J.A.,
Gravitation, San Francisco: Freeman, 1973.

11. Particle Data Group (Barnett, R.M. et at.),
Phys. Rev. D: Part. Fields, 1996, vol. 54, p. 1.

12. Freedman, W.L. et al., Nature, 1994, vol. 371, p. 757.

13. Peebles, P.J.E., The Large-Scale Structure of the Universe,
Princeton: Princeton Univ. Press, 1980.

14. Smoot, J.F. and Scott, D., Eur. Phvs. J. C: Part. Fields, 1998,
vol. 3, p. 127.

\end{document}